# Non-Reference Quality Monitoring of Digital Images using Gradient Statistics and Feedforward Neural Networks

Nisar Ahmed[1], Hafiz Muhammad Shahzad Asif[2], Hassan Khalid[3]

[1,3] Space and Upper Atmosphere Research Commission, Pakistan.
[1,2] Department of Computer Science and Engineering, University of Engineering and Technology, Lahore, Pakistan.
[1]nisarahmedrana@yahoo.com, [2]shehzad@uet.edu.pk, [3]engr_hassan09@ymail.com

*Abstract*— Digital images contain a lot of redundancies, therefore, compressions are applied to reduce the image size without the loss of reasonable image quality. Same become more prominent in the case of videos which contains image sequences and higher compression ratios are achieved in low throughput networks. Assessment of quality of images in such scenarios become of particular interest. Subjective evaluation in most of the scenarios become infeasible so objective evaluation is preferred. Among the three objective quality measures, full-reference and reduced-reference methods require an original image in some form to calculate the quality score which is not feasible in scenarios such as broadcasting or IP video. Therefore, a non-reference quality metric is proposed to assess the quality of digital images which calculates luminance and multiscale gradient statistics along with mean subtracted contrast normalized products as features to train a Feedforward Neural Network with Scaled Conjugate Gradient. The trained network has provided good regression and $R^2$ measures and further testing on LIVE Image Quality Assessment database release-2 has shown promising results. Pearson, Kendall and Spearman's correlation is calculated between predicted and actual quality scores and their results are comparable to the state-of-the-art systems. Moreover, the proposed metric is computationally faster than its counterparts and can be used for quality assessment of image sequences.

## I. Introduction

In most of the applications, quality of an image is momentous. There are a number of techniques currently available for the evaluation of an image quality. There are broadly two classes of such evaluation techniques, the first class requires reference images while the second class does not require any reference images. In maximum cases, reference images are not available so image quality evaluation without reference images is of significant importance. Perceptual quality of an image is more noteworthy than the actual quality of an image as the perception of the quality of an image by the observer is of more significance. There are broadly two classes of perceptual image quality assessment (IQA) techniques currently available: subjective evaluation by the humans and objective evaluation through the algorithms in order to parodist the subjective evaluation [1-3]. Objective assessment techniques are less expensive, less time consuming they are of significant importance [1].There are three categorize of IQA models.

Non-reference (NR) approaches, full-reference (FR) approaches and reduced-reference (RR) approaches [2]. The non-reference IQA in their early stages assumed that the quality of an image is affected by a certain particular type of distortions i.e. blackness, ringing, blurriness, and compression [1]. The early techniques, therefore, utilize the presumed distortion models and so have limited scope of applications [1]. In order to resolve the problem of limitation in the NR-IQA, researchers have developed a new method of perceptual image quality assessment known as blind image quality assessment (BIQA). In BIQA, there is no need to have the presumed distortion models; the only information needed is the image for which quality evaluation is required. Most of the currently existing BIQA methods are dependent on the opinion, they are trained on the subjective scores consisting images dataset [1]. The different type of methods for BIQA are [1-10]. In [5], Saad et al. proposed a method for BIQA based on the DCT extracted structure and contrast features. They introduced the BLIINDS (BLind Image Integrity Notator using DCT Statistics) index, it is a NR-IQA model which use probabilistic prediction model trained on a small dataset. The model mainly relies on covariance and mean of the training data for prediction.

In [4], Saad et al. proposed a BIQA method, based on natural scene statistics (NSS) model of discrete cosine transform (DCT) coefficients. They name the method as BLIINDS-II which is an extension of their previous work which use some additional natural scene statistics and introduction of Bayesian inference in prediction. In [8], Ye et al. proposed an algorithm that uses the patches of the raw image as there local descriptors and for encoding soft-assignment. They have tested CORNIA (Codebook Representation for No-Reference Image Assessment) on the LIVE database and shown that its statistical performance is better than the full-reference quality evaluation. In [10], Mittal et al. proposed completely blind image quality analyzer which uses "opinion aware" BIQA method. The proposed IQA model which is called as Natural Image Quality Evaluator (NIQE) is based on the creation of a 'quality aware' statistical features collection that is based on natural scene statistic (NSS) model. In [1], Xue et al. proposes a novel BIQA method based on joint statistics of local image gradient magnitudes and the Laplacian of Gaussian image responses. In [2], Zhang et al. proposed a completely blind image quality evaluator using Statistics of Normalized

Luminance, Statistics of MSCN Products, Gradient Statistics, Statistics of Log-Gabor Filter Responses and Statistics of Colors. Their approach is feature enriched and take slightly more time in computation of feature set. In [3], Liu et al. proposed BIQA by relative gradient statistics and adaboosting neural networks. Their approach calculates gradient magnitude and other relative gradient statistics for prediction of image quality. The feature extraction is performed at two different image scales to capture deeper granularities. Finally, the prediction is made by using adaboosting neural network which train some weak learners and assign weight to each learner based on its prediction accuracy. The final prediction is determined based on prediction of each neural network multiplied with their weights which makes the algorithm comparatively slow in training and prediction. In [7], Ghadiyaram et al. proposed a technique of Perceptual image Quality Prediction based on Authentically Distorted Images by using Natural scene statistics (NSS) model a Bag of Features approach. There are few other bling image quality assessment approaches which utilize different types of natural scene statistics and perform regression modeling for prediction of objective image quality.

In our approach, we have used gradient and relative gradient statistics which appear to be a strong predictor of human subjective score. The original approach they have taken these statistics at two image scales whereas we have extracted these statistics at deeper scales and realized that gradient statistics at three scales provide good prediction and further scaling doesn't add much to the feature space. Moreover, to expand the feature set we further introduced statistics of image luminance and Mean Subtracted and Contrast Normalized (MSCN) coefficient products and found that it is pretty helpful in increasing the predictive accuracy of blind IQA. Finally, a three layered feedforward neural network with scaled conjugate gradient training is used on GPU to train and validate the NN with different parameters and a fine tuned network is selected which performed good on the benchmark dataset of LIVE IQA database and its results are comparable with the other approaches and better then the approach which we tried to improve.

## II. METHODOLOGY

It is observed that Natural Scene Statistics (NSS) are an exceptional indicator of the perceptive quality of natural images [3-5, 9-13]. As a result, NSS based models are being widely used in Image Quality Assessment (IQA). Gradient statistics in combination to other natural scene statistics can be used in an optimal fashion to design an IQA system which can provide a good perceptual quality prediction. It is demonstrated in previous studies that distortion in image quality can be considered from features of local structural contrast, gradient, luminance and color statistics [1, 14, 15]. In our study, we have combined an appropriate set of gradient statistics with Laplacian features and normalized luminance statistics. The method discussed is a Non-reference, opinion aware regression model which maps the features to mean opinion score acquired from subjective quality scores. The study is focused on estimation of quality of digital images which are distorted with commonly occurring distortion in digital images.

### A. Gradient Statistics

The image gradient delivers extensive information regarding the structural properties of the image so it is extensively used in IQA in different forms. Different gradient based statistics are used by various researchers and their approaches have provided varying information [1, 3, 6, 12, 14-17]. However, most gradient based approaches are either full-reference (FR) or reduced-reference (RR) and some are non-reference (NR) IQA approaches are specific to distortion type except few exceptions [1-3, 18]. Owing to the success of these approaches in predicting bling image quality using gradient statistics there seems to be a strong bond between image quality and its gradient. We have relied on the gradient statistics extraction model of [3] which relies on Gradient Magnitude (GM), Relative gradient Orientation (RO), and Relative gradient Magnitude (RM). The estimated image gradient can be computed from the Equation 1.

$$|\nabla I(i,j)| = \sqrt{I_x^2(i,j) + I_y^2(i,j)} \quad \text{Eq. 1}$$

Where $I_x$ and $I_y$ are values of directional derivatives in the horizontal $x$ and vertical $y$ directions corresponding to sample directions $i$ and $j$.

The distortions in image exert a deep influence on gradient magnitude particularly when the image face blurring distortion [3]. Whereas in other types of distortions, the change in magnitude may not be much evident and though the image is still visually different from the original image. As an example, JPEG compression produces significant distortion which is visually apparent and still the gradient magnitude histograms of the perfect and distorted image are not significantly different making gradient magnitude alone less valuable feature. The gradient magnitude delivers information regarding brightness variations only whereas the human visual cortex is highly sensitive to orientation information as well [3]. As the distortions in image result in modification of local image anisotropies, the use of gradient orientation along with gradient magnitude will be beneficial in the prediction of image quality. Similarly, relative gradient magnitude delivers much information regarding the local structures in the image. The orientation of gradient delivers information additional to gradient magnitude which results in improved IQA [3]. The estimated gradient orientation can be given by Equation 2.

$$< \nabla I(i,j) = \arctan \frac{I_y(i,j)}{I_x(i,j)} \quad \text{Eq. 2}$$

The orientation of gradient can be measured in an absolute manner against the reference frame of spatial coordinate system of the image or it could be measured relatively against the average image values.

The later model is more relevant to IQA as relative gradient orientation captures local degradations of image structure. Accordingly, three gradient maps, GM, RM and RO are computed from $I_x$ and $I_y$ to describe the image gradient behavior with respect to perceptive quality over blocks of $M \times N$ dimensions. The GM is calculated using equation 1 whereas the RO can be calculated using Equation 3.

$$< \nabla I(i,j)_{RO} = < \nabla I(i,j) - < \nabla I(i,j)_{AVE} \qquad \text{Eq. 3}$$

And the average local orientation can be computed using Equation 4.

$$< \nabla I(i,j)_{AVE} = \arctan \frac{I_y(i,j)_{AVE}}{I_x(i,j)_{AVE}} \qquad \text{Eq. 4}$$

Where

$$I_x(i,j)_{AVE} = \frac{1}{MN} \sum_{(m,n) \in W} I_x(i-m, j-n)$$

and

$$I_y(i,j)_{AVE} = \frac{1}{MN} \sum_{(m,n) \in W} I_y(i-m, j-n)$$

Where $W$ defines a local neighborhood over which the derivative is taken. In this implementation a neighborhood of $3 \times 3$ is taken having $M = N = 3$, for which values of $W$ coordinates are $(-1,-1), (-1,1), (1,-1), (-1,0), (0,-1), (0,0), (0,1), (1,0)$ and $(1,1)$. Similarly, RM is given by Equation 5:

$$|\nabla I(i,j)|_{RM} = \sqrt{\left(I_x(i,j) - I_x(i,j)_{AVE}\right)^2 + \left(I_y(i,j) - I_y(i,j)_{AVE}\right)^2} \qquad \text{Eq. 5}$$

The directional gradient components $I_x$ and $I_y$ are computed using Gaussian partial derivative filters. In comparison to other filters such as Prewitt and Sobel, the Gaussian partial derivative can be treated as smoothed edge filter which is beneficial for computing RO and RM. Equation 6 can be used for computation of Gaussian partial derivative:

$$\nabla_\gamma G(x,y,\sigma) = -\frac{\gamma}{2\pi\sigma^4} \exp\left(-\frac{x^2+y^2}{2\sigma^2}\right) \qquad \text{Eq. 6}$$

As mentioned in [3] variations in statistical distributions of gradient and relative gradient measures due to distortions can be used for quantification of perceived distortion in images. These gradient distributions are characterized by using histogram of variance as described in [19]. Though all the characteristics of histograms cannot be expressed through variances but still, they signify most notable characteristics. The variance of a histogram $\nabla(x)$ can be defined by the Equation 7:

$$Var(\nabla) = \sum_x (\nabla(x) - \bar{\nabla})^2 \qquad \text{Eq. 7}$$

Resultantly we obtain a three-dimensional feature vector:

$$Feat\_Vec = [V_{GM} \quad V_{RO} \quad V_{RM}]$$

To take the multistate characteristics of natural images for distortion and visual perception [4, 20], we have downsampled the image to $\frac{1}{2}$ and $\frac{1}{4}$ and then computed the same set of three features producing a total of 9 features. It is to note that further scaling to extract more feature didn't result in a significant gain in predictive accuracy.

### B. Statistics of Standardized Luminance

It was highlighted in [19] that the standardized luminance of an image $I$ adapt to a Gaussian distribution. The standardization of the natural image can be done using equation 8.

$$\bar{I}(i,j) = \frac{I(i,j) - \mu(i,j)}{\sigma(i,j) + 1} \qquad \text{Eq. 8}$$

Whereas i and j are spatial coordinates of pixels and

$$\mu(i,j) = \sum_{k=-K}^{K} \sum_{l=-L}^{L} \omega_{k,l} I(i+k, j+l)$$

$$\sigma(i,j) = \sqrt{\sum_{k=-K}^{K} \sum_{l=-L}^{L} \omega_{k,l} (I(i+k, j+l) - \mu(i,j))^2}$$

are the mean and contrast of the local image. The structural distortion is characterized from local MSCN coefficients and from products of adjacent pairs distribution of these coefficients. These product coefficients $\bar{I}(i,j)$ follow a unit Gaussian distribution on digital images which are not perceptually distorted [19]. This Gaussian model is disrupted when the images undergo some distortion. Measurement of deviance of $\bar{I}(i,j)$ from Gaussian model indicate the level of distortion. According to suggestion in [9] & [10], a zero-mean Generalized Gaussian Distribution (GGD) model can generally channelize the dispersal in the incidence of distortions. The density function of GGD can be given by Equation 9.

$$g(x; \alpha, \beta) = \frac{\alpha}{2\beta \Gamma\left(\frac{1}{\alpha}\right)} \exp\left(-\left(\frac{|x|}{\beta}\right)^\alpha\right) \qquad \text{Eq. 9}$$

Where $\Gamma(.)$ is the gamma function

$$\Gamma(x) = \int_0^\infty t^{x-1} e^{-t} dt, \quad x > 0$$

The parameters $\alpha$ and $\beta$ in equation 2 are effective features which can be estimated using the moment matching approach with reliance. [21].

### C. Statistics of MSCN Products

Image quality information can also be captured through products of adjacent pairs of MSCN coefficients as described in [10] and [9]. The distribution of products pairs particularly $\bar{I}(i,j) \bar{I}(i,j+1)$, $\bar{I}(i,j) \bar{I}(i+1,j+1)$, $\bar{I}(i,j) \bar{I}(i+1,j-1)$ and $\bar{I}(i,j) \bar{I}(i+1,j)$ is calculated to describe image quality. These products are demonstrated on perfect and distorted images following a zero mode asymmetric GGD (AGGD) [22]. AGGD can be calculated using Equation 10.

$$g_\alpha(x; \gamma, \beta_l, \beta_r) = \begin{cases} \frac{\gamma}{(\beta_l + \beta_r)\Gamma\left(\frac{1}{\gamma}\right)} \exp\left(-\left(\frac{-x}{\beta_l}\right)^\gamma\right), & x \le 0 \\ \frac{\gamma}{(\beta_l + \beta_r)\Gamma\left(\frac{1}{\gamma}\right)} \exp\left(-\left(\frac{x}{\beta_r}\right)^\gamma\right), & x > 0 \end{cases} \qquad \text{Eq. 10}$$

The mean of the AGGD is given by Equation 11:

$$\eta = (\beta_r - \beta_l) \frac{\Gamma\left(\frac{2}{\gamma}\right)}{\Gamma\left(\frac{1}{\gamma}\right)} \qquad \text{Eq. 11}$$

These four parameters ($\gamma$, $\beta_l$, $\beta_r$, $\eta$) are also effective features which can describe image quality. These four features can be calculated in four orientations as mentioned in earlier paragraph and a set of features can be acquired.

### D. Feedforward Neural Network

Artificial Neural Networks are inspired from the organization and decision-making process of the human brain [23]. These networks can be used for curve fitting, pattern recognition and clustering of data. They are preferably used when the problem system is complex and its underlying process is not clear. Feedforward (FF) neural network takes some input and provide and output based on its structure, weights, and activation functions and the output is net fed back to the network to form a cycle. A two layer FF network is easy to train by calculating the error between the predicted and expected values and adjusting the weights and biases. Multilayer network, on the other hand, cannot be trained in this way and they required backpropagation of error to neurons of earlier layers to adjust weights. There are several training algorithms (backpropagation being the first attempt) to train a multilayer FF neural networks as they are necessary to solve complex and nonlinear problems. Successful development of a FF neural network involves three main stages:
    a. Definition of network topology
    b. Selection of a suitable training algorithm
    c. Activation function for output and intermediate layers

Fundamentally a NN contains an input layer, one or more connected hidden layers and an output layer. The learning involves adjustment of weights and biases of interconnections between the layers. One the completion of the learning process, the network produces a suitable output at the output layers. The learning process may be supervised or unsupervised depending on the requirements. In our scenario, supervised learning is adopted in which network calculates the error from desired output to adjust the weights and biases.

### E. Proposed Approach

In our problem, we have calculated two set of features. The first one is gradient statistics containing GM, RO and RM calculated at three different scales forming up a feature vector of nine features. The second one is a set of 18 features, two of them are calculated from normalized luminance and the sixteen of them are calculated from MSCN products. The two feature sets are concatenated to form a final feature vector of 27 features. Prior to initialization of training process, we normalized these feature using the formula in equation 12.

$$feat_{new} = \frac{x - \bar{x}}{\sigma} \qquad \text{Eq. 12}$$

Where $x$ is initial feature value, $\bar{x}$ is the mean of feature values and $\sigma$ is the standard deviation.

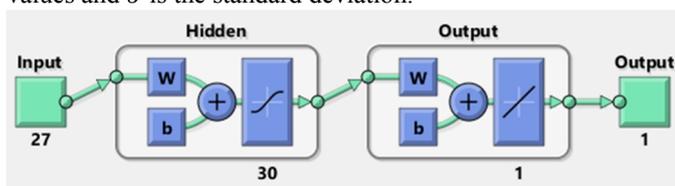

*Fig. 1. Topology of neural network*

The neural network topology used for experimentations is provided in Figure 1. The input layer of the network contains 27 neurons which are connected through weights, hidden layer contains 30 neurons and the output layer contains one neuron. The number of neurons in the hidden layer is varied from 10 to 40 and the layers were increased up to three but the best performance was achieved with 30 neurons with single hidden layer. The training of the algorithm is done using Scaled Conjugate Gradient (SCG) which is a fast training algorithm and can operate using a GPU. As several iterations were required to select the parameters of the network and to fine tune the neural network a faster version of training algorithm with the capability to operate using GPU was preferred. Training of SCG automatically stops when the generalization of the network stops on validation samples. Table 1 provides the parameters of the final network.

*TABLE 1 Training Neural Network Parameters*

| Parameter | Value |
| --- | --- |
| Training Data | 70% |
| Validation Data | 15% |
| Testing Data | 15% |
| Performance Measure | Mean Square Error |
| Max Epochs | 2000 |
| Number of weights | 871 |

The network performance curves provided in Figure 2 shows the best validation performance with a MSE of 32.1342 is achieved at epoch 586. The regression lines for training, validation and testing data are provided in Figure 3. Figure 4 provides the gradient values after each epoch. Finally, the Figure 5 provides the error histogram which is nearly centered at zero value and also show spreading of error beyond this value.

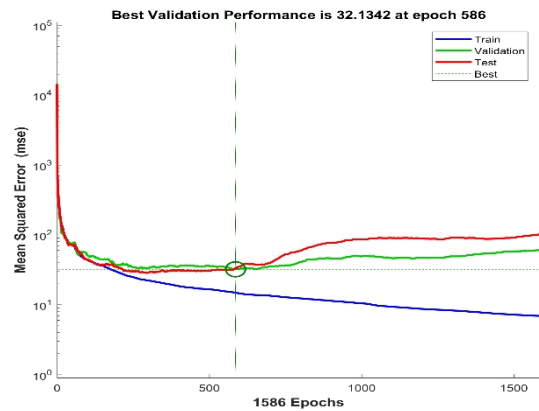

*Fig. 2. Performance curves with best validation performance*

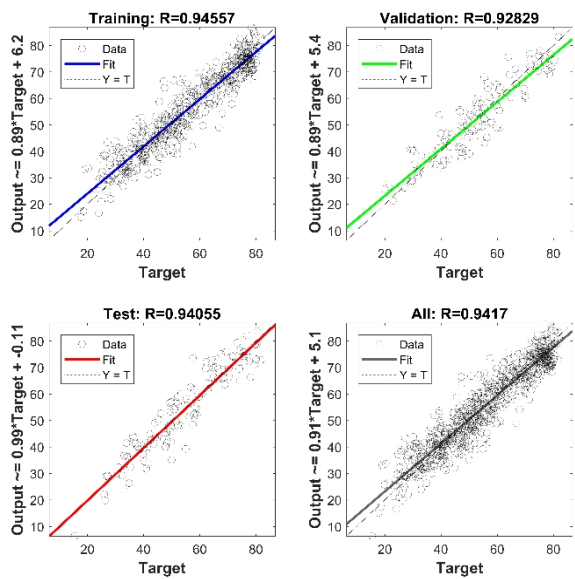

*Fig. 3. Regression of training, validation and testing scenarios*

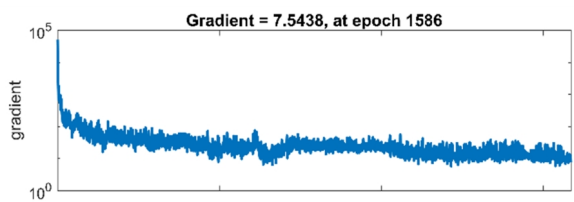

*Fig. 4. Decreasing gradient trend*

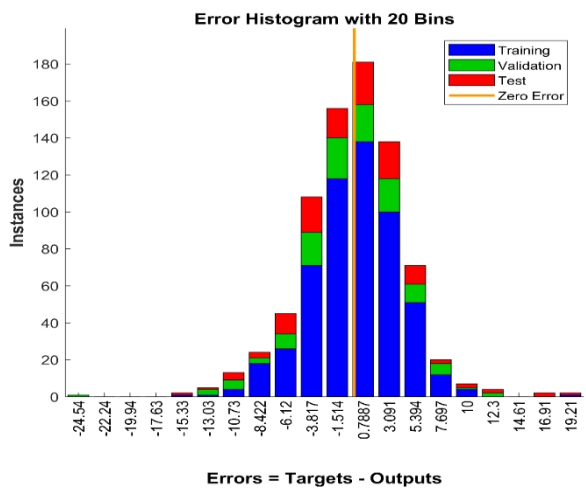

*Fig. 5. Error Histogram of training, validation, and testing*

### III. RESULTS

LIVE database release-2 is used for testing of the proposed network. There are five different types of distortion in the image database. The database is generated from 29 perfect images and then different simulated distortions are added in these images and their subjective quality evaluation is performed to generate mean opinion score (MOS) and difference MOS. The subjective evaluation is done in same environmental conditions and on average 30 persons gave their subjective evaluation experience for each image. Table 2 provides the five different simulated distortions and the number of images produced from this distortion. These are the most common type of distortions faced by digital images during storage and communication. JPEG and JPEG 2000 are the most widely used lossy compression schemes and most of natural images have to be compressed by one of these. Rayleigh fast fading and white noise is simulated on the images as it is the most widely faced distortion during image communication especially in wireless environment. Gaussian blur on the other hand is added due to number of reason and handshake, lens blur or motion effect are the most common type of blurring which is demonstrated through the use of Gaussian blur.

*TABLE 2 Distortion Type and number of images produced from it*

| Distortion Type | Number of Images |
|---|---|
| JPEG 2000 compression | 227 |
| JPEG compression | 233 |
| White Noise | 174 |
| Gaussian Blur | 174 |
| Fast Fading | 174 |

The trained network is used to predict the objective image quality based on the training and the results for combined image database as well as each distortion type are presented in Table 3. Three different correlation test are run on predicted and MOS quality of the image and their results are provided. It is evident that the predicted image quality is very similar to the MOS provided during training. It is persistent to mention that algorithm performed exceptionally good on images distorted with white noise whereas the prediction in distorted images through fast fading is somewhat below its counterparts. Some more features could be explored and combined in an optimized form to increase the prediction accuracy of images distorted through fast fading.

*TABLE 3 Correlation Test of LIVE dataset*

|  | JP2K | JPEG | WN | G. Blur | FF | All |
|---|---|---|---|---|---|---|
| **Pearson** | 0.9195 | 0.9215 | 0.9328 | 0.9127 | 0.8639 | 0.9106 |
| **Kendall** | 0.7882 | 0.7569 | 0.8408 | 0.7897 | 0.7413 | 0.7804 |
| **Spearman** | 0.9463 | 0.9209 | 0.9592 | 0.9383 | 0.9108 | 0.9343 |

### IV. CONCLUSION

An efficient blind image quality assessment system is proposed in this study which predicts the quality of a digital image based on some training algorithm. The training is done based on MOS which is obtained from human observers by showing them perfect and distorted digital images. The distorted images are generated through several simulated distortions which are commonly faced by digital images during storage, communication or compression. The system work by extracting a set of features which is carefully crafted and optimized for the existing features presented by different researchers. The gradient statistics is extracted at three different scales as opposed to original approach which yielded better performance. The increase in performance in multiscale model is due the fact that human perceive image details differently at different scales. Then a feedforward neural network is constructed and trained by tuning its different parameters for best performance. Different performance parameters of tuned neural network are provided for demonstration. Finally, the network is tested on whole image database as well as on individual distorted images and then the correlation is calculated between predicted quality and MOS which show comparable performance with state-of-the-art algorithms.

*A. Future Recommendation*

The future direction of work includes exploration of more features, especially from transform domain to better represent the image quality which correlate with human visual perception. The optimized combination of features which are fast to extract is an essence to a successful image quality analyser. Some feature selection measure could be used to rank the feature which perform better than other in prediction of desired quality. Moreover, there are numerous other database for the same task and the proposed approach could be testing on them to check if it conforms to the requirements. Video quality assessment could also be an extension of this work as emphasis was paid to faster processing and prediction of IQA score so it could be used in real-time video quality assessment may be with the help of some GPU acceleration.